\documentclass[lettersize,journal]{IEEEtran}
\usepackage{amsmath,amsfonts}
\usepackage{algorithmic}
\usepackage{algorithm}
\usepackage{array}
\usepackage[caption=false,font=normalsize,labelfont=sf,textfont=sf]{subfig}
\usepackage{textcomp}
\usepackage{stfloats}
\usepackage{url}
\usepackage{verbatim}
\usepackage{graphicx}
\usepackage{cite}
\begin{document}

\title{CRS-LLM: Cooperative Beam Prediction with a GPT-Style Backbone and Switch-Gated Fusion}

\author{
    Fangzhi~Li,
    Cunhua~Pan,~\IEEEmembership{Senior~Member,~IEEE},
    Hong~Ren,~\IEEEmembership{Member,~IEEE},\\
    Dongming~Wang,
    and~Jiangzhou~Wang,~\IEEEmembership{Fellow,~IEEE}
        
\thanks{F. Li, C. Pan, and H. Ren are with the National Mobile Communications Research Laboratory, Southeast University, Nanjing 211189, China. \\

D. Wang and J. Wang are with the National Mobile Communications Research Laboratory, Southeast University, Nanjing 211189, China, and also with Purple Mountain Laboratories, Nanjing 211111, China.}}

\markboth{ }
{CRS-LLM: Cooperative Beam Prediction with a GPT-Style Backbone and Switch-Gated Fusion}

\maketitle

\begin{abstract}
Millimeter-wave (mmWave) communication relies on highly directional beamforming, but fast mobility, blockage, and rapid geometry changes in vehicle-to-everything (V2X) scenarios make reliable beam tracking highly challenging. In cooperative systems with multiple base stations (BSs), conventional hierarchical methods usually perform BS selection and beam selection separately. This two-stage process can suffer from error propagation when the beam state changes abruptly. To address this problem, this paper proposes Cooperative Radio Sensing with Large Language Models (CRS-LLM), a cooperative beam prediction framework with a Generative Pre-trained Transformer (GPT)-style backbone and switch-gated fusion for next-step joint BS--beam prediction. The method formulates beam tracking as a single classification problem over a joint BS--beam space, which avoids the error propagation caused by cascaded decisions. To connect wireless channel state information (CSI) with large language models (LLMs), we design a dual-view CSI tokenizer that uses both frequency-domain and delay-domain channel representations. A lightweight convolutional neural network (CNN) front-end and a temporal tokenization module are then applied. A truncated pre-trained GPT-style backbone is adopted for temporal modeling with parameter-efficient adaptation. A transition-aware switch-gated predictor further combines a stable branch, a residual flip branch, and a low-rank transition prior to handle both smooth beam evolution and abrupt beam changes. Simulation results show that CRS-LLM outperforms the considered baselines, including CSI-Transformer, Hierarchical BS--Beam, and representative CNN- and recurrent-neural-network-based methods, in Top-1 accuracy and normalized beam gain across different signal-to-noise ratio (SNR) conditions, while also showing strong few-shot performance and promising zero-shot transferability to unseen propagation scenarios.
\end{abstract}


\section{Introduction}

\IEEEPARstart{M}{illimeter-wave} (mmWave) communication and massive multiple-input multiple-output (MIMO) are widely viewed as important enablers for 5G-Advanced and 6G because they offer large bandwidth and high spatial multiplexing capability \cite{saad2020vision6g,zhang2019vtn6g,zong2019vtn6gtech,yang2019mnet6gvision,tataria2021jproc6g}. However, operation at these frequencies depends heavily on directional beamforming to overcome severe path loss and blockage sensitivity \cite{busari2018mmwave_mimo_survey}. This issue is particularly critical in urban vehicle-to-everything (V2X) scenarios, where mobility, frequent blockage, and rapidly changing geometry make stable beam alignment difficult \cite{heng2021mcom_beamchallenges,tan2021jcs_isac6g}.

In such environments, beam management must continuously support beam acquisition, refinement, tracking, and recovery. Existing studies have shown that this process is necessary in mmWave systems, but it can also introduce high training and signaling overhead, especially when users move quickly or the channel changes abruptly \cite{giordani2019comst_beam_mgmt_tutorial,kalamkar2022twc_beam_mgmt_sg,ma2023iwc_dl_beam_mgmt,xue2024comst_beam_mgmt_survey,khan2023access_ml_beam_mgmt_survey}. Therefore, reducing beam management overhead while maintaining beam quality has become a key problem for practical mmWave and near-terahertz (THz) systems.

Predictive beamforming is an effective approach to reducing this overhead. Instead of repeatedly sweeping candidate beams, the system infers the future beam state from historical observations and side information. Prior studies have demonstrated the value of this idea in sensing-assisted vehicular links, Bayesian predictive beamforming, and dynamic vehicle-to-infrastructure (V2I) scenarios \cite{liu2020twc_radar_assisted_pb,yuan2021twc_bayesian_pb,du2023twc_isac_v2i}. Related learning-based predictive beamforming methods have also been reported for unmanned aerial vehicle (UAV) communications and mobile mmWave tracking \cite{yuan2020wcl_uav_jittering,liu2021wcl_uav_location,lim2021tcom_beam_tracking,liu2021isac_predictive_beamforming}. These results suggested that prediction can greatly reduce tracking cost, but they also showed that the problem becomes much harder when mobility, blockage, and handover happen together.

This challenge is even more pronounced in cooperative systems with multiple base stations (BSs). Coordinated beamforming across multiple BSs can improve robustness in highly mobile mmWave systems \cite{alkhateeb2018access_coordinated_bf}, and related ideas have also been studied for beam management and interference coordination in dense deployments \cite{zhou2019tvt_dl_beam_mgmt_interference}. More recently, hierarchical beam alignment has shown that structure-aware prediction can reduce search complexity \cite{yang2024twc_hierarchical_beam_alignment}. Meanwhile, the broader development of cell-free and distributed massive MIMO systems has highlighted the value of BS cooperation for coverage continuity and macro-diversity \cite{ngo2017twc_cellfree,interdonato2019jwcn_cellfree}. These trends suggest the need for beam prediction methods that can exploit joint information from several BSs rather than relying on one BS alone.

However, many existing designs still decompose the decision into two stages, such as first choosing the serving BS and then selecting the beam within that BS. This decomposition is intuitive, but it can suffer from error propagation. Once the BS decision is wrong, the beam decision is made within the wrong search space. In fast V2X scenarios, this weakness becomes more severe because the best BS--beam pair may change abruptly under blockage, turns, or handover. For this reason, a flat joint BS--beam formulation is often more suitable for cooperative prediction than a cascaded structure.

Another difficulty lies in the form of the input itself. Cooperative wideband channel state information (CSI) is high-dimensional, complex-valued, and jointly structured over space, frequency, and time. More broadly, the literature on deep learning for the physical layer has shown both the promise and the limitations of purely data-driven wireless designs \cite{huang2020mwc_dl_phys_layer_5g,he2019iwc_model_driven_dl,oshea2017intro_dl_phy,qin2018dl_phy}. For example, CSI feedback networks have demonstrated that neural models can learn useful channel structure from compact representations \cite{wen2018wcl_csinetwork}, while recent work on personalized federated CSI learning has further emphasized the difficulty of data heterogeneity and cross-device generalization \cite{cui2023personalized_federated_csi}. These observations suggest that cooperative beam prediction needs not only a strong backbone, but also an input representation that preserves the structure of wireless data.

A related issue is that real beam evolution is not uniform over time. In many slots, the preferred BS--beam pair evolves smoothly and is strongly correlated with the recent trajectory. In other slots, however, the preferred pair can jump suddenly because of blockage, fast motion, or inter-BS switching. A single classifier trained with one shared objective may perform well on the common smooth cases, but still remain weak on rarer yet practically important transition events. These two regimes place different demands on the predictor and motivate an output design that can adapt to both continuation-dominated and transition-dominated beam evolution.

Recently, transformer-based models have shown strong capability in sequence modeling, especially when long-range dependence and variable temporal patterns are important \cite{wen2023ijcai_ts_transformers}. At the same time, recent discussions on large artificial intelligence (AI) models for telecom and 6G have suggested that foundation-model-style backbones may become useful for wireless tasks when combined with suitable domain adaptation and representation design \cite{bariah2024mcom_large_genai_telecom,chen2024mwc_big_ai_models_6g}. This motivates the use of a pre-trained Generative Pre-trained Transformer (GPT)-style sequence backbone for cooperative beam prediction. However, such a backbone cannot be applied directly. Wireless CSI is very different from natural language tokens, so the model will not work well unless this gap is properly handled.

To address these issues, we propose Cooperative Radio Sensing with Large Language Models (CRS-LLM), a cooperative next-step joint BS--beam prediction framework for dynamic mmWave V2X scenarios. In this paper, ``cooperative radio sensing'' refers to using cooperative radio observations as implicit cues of geometry evolution, blockage variation, and path-transition regularity, rather than introducing an explicit sensing task or a dedicated sensing module. The proposed method uses a flat joint label space so that intra-BS beam reselection and inter-BS handover are handled in a unified manner. This avoids the hard dependency of beam prediction on an earlier BS classification stage and makes the model better suited to abrupt cooperative switching events.

To connect cooperative CSI with a pre-trained sequence backbone, we design a dual-view CSI tokenizer, which jointly exploits frequency-domain and delay-domain channel representations, allowing the model to preserve wideband structure while exposing multipath sparsity more clearly. A lightweight convolutional neural network (CNN) front-end extracts compact local features, and a temporal tokenization process converts them into a sequence that can be processed by a truncated GPT-style backbone. In this way, the model can use temporal context from multiple past slots while keeping the radio-domain structure visible to the backbone.

On top of this representation, we further introduce a transition-aware switch-gated prediction head. Instead of using a single output branch, the predictor combines a stable branch, a residual flip branch, and a low-rank transition prior. The stable branch focuses on cases with stable continuation, whereas the residual branch emphasizes abrupt label changes. A soft gate then fuses the two outputs adaptively according to the current latent representation. This design is intended to better match the coexistence of stable evolution and abrupt transitions in practical V2X beam sequences.

The main contributions of this paper are summarized as follows:
\begin{itemize}
    \item A unified joint BS--beam prediction framework: We formulate cooperative beam tracking as a flat classification problem over a joint BS--beam space. This avoids the error propagation of hierarchical BS-first and beam-second prediction and handles intra-BS switching and inter-BS handover in one stage.

    \item A dual-view CSI tokenization method for LLM-based modeling: We design a tokenizer that organizes cooperative CSI into frequency-domain and delay-domain views, followed by a lightweight CNN front-end and structural embedding, thereby bridging the gap between wireless CSI tensors and the token sequence required by a GPT-style backbone.

    \item A transition-aware switch-gated prediction head: We introduce a prediction head that combines a stable branch, a residual flip branch, and a low-rank transition prior, so that smooth beam continuation and abrupt beam changes can be modeled in a more targeted way.

    \item Strong performance and adaptation ability: Simulation results show that the proposed CRS-LLM improves Top-1 accuracy, which measures whether the highest-scored predicted BS--beam label is correct, and normalized beam gain over strong baselines, including CSI-Transformer, Hierarchical BS--Beam, and representative CNN- and recurrent-neural-network-based methods, under different signal-to-noise ratio (SNR) conditions, while also showing favorable few-shot adaptation and cross-scenario transfer performance.
\end{itemize}

\textit{Notation:} Throughout this paper, boldface lowercase and uppercase letters denote vectors and matrices (or higher-dimensional tensors), respectively. Calligraphic letters denote sets. The superscripts $(\cdot)^T$ and $(\cdot)^H$ denote transpose and conjugate transpose, respectively. $\mathbb{R}$ and $\mathbb{C}$ represent the real and complex spaces. $\|\cdot\|_2$ and $\|\cdot\|_F$ denote the $\ell_2$-norm and Frobenius norm. For a vector $\mathbf{a}$, $\mathbf{a}[i]$ denotes its $i$-th element. For a matrix $\mathbf{A}$, $\mathbf{A}[i,j]$ denotes the element in the $i$-th row and $j$-th column, and $\mathbf{A}[:,i]$ denotes the $i$-th column. Multi-dimensional slicing is denoted by colons, e.g., $\mathbf{X}[\tau,b,v,:,:,:]$ denotes a slice of tensor $\mathbf{X}$ over the last three dimensions.

\section{System Model}

\subsection{Network Architecture and Cooperative Beam Tracking Scenario}

Consider a cooperative mmWave/orthogonal frequency-division multiplexing (OFDM) \cite{Chunk2009ZhuI, Chunk2009ZhuII} beam tracking system in a dynamic urban V2X scenario. The network contains $N_{\mathrm{BS}}$ geographically distributed BSs, denoted by
\begin{equation}
\mathcal{B}=\{1,2,\ldots,N_{\mathrm{BS}}\},
\end{equation}
which jointly serve a mobile user equipment (UE). The BSs are connected to an edge controller through low-latency backhaul links, so that the observations collected at different BSs can be aggregated and used for centralized prediction. This architecture reflects a practical cooperative beam-tracking pipeline in which geographically distributed BSs provide partially complementary views of the same UE trajectory and propagation condition. As a result, the predictor is not restricted to the instantaneous visibility of one serving BS, but can exploit cross-BS diversity to improve robustness when local blockage or geometry change weakens the current link.

\begin{figure}[h]
    \centering
    \includegraphics[width=\linewidth]{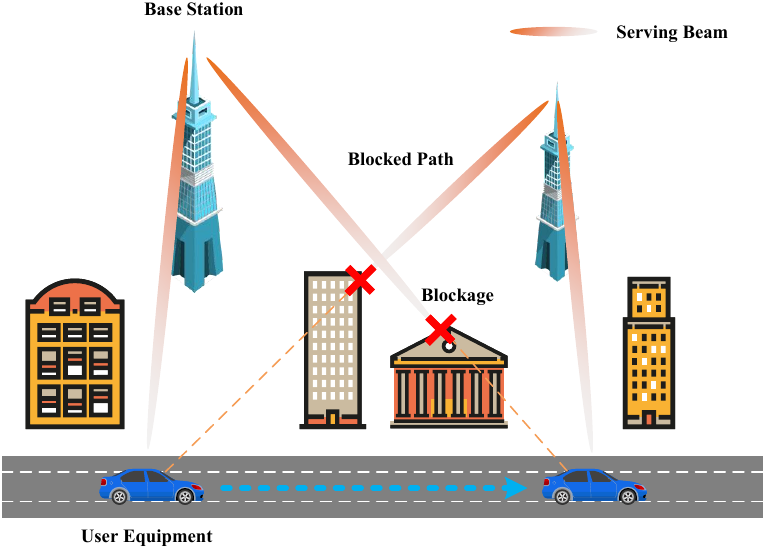}
    \caption{Cooperative multi-BS beam tracking scenario. Because of mobility and blockage, the current serving path may become unavailable, which makes both inter-BS and intra-BS beam reselection necessary.}
    \label{fig:physical_scenario}
\end{figure}

Fig.~\ref{fig:physical_scenario} illustrates the considered setting. Because of mobility, blockage, and rapid geometry variation, the currently preferred BS--beam pair may not remain optimal in the next slot. The future change may occur within the current BS through beam reselection, or across BSs through inter-BS handover. For this reason, the prediction task is formulated directly in the joint BS--beam space rather than decomposed into separate BS selection and beam selection stages, thereby avoiding error propagation between cascaded decisions.

\subsection{Wideband Multi-BS Channel Representation}

Let $t$ denote the slot index and let $t_s=t\Delta t$ denote the physical time of slot $t$, where $\Delta t$ is the slot duration. Each BS is equipped with a planar antenna array, and the effective number of BS antenna ports is denoted by $N_p$. In the simulations, a dual-polarized $4\times4$ BS array is used, which gives $N_p=32$ effective ports. The UE is equipped with a single antenna. Therefore, for BS $b\in\mathcal{B}$, the frequency-domain downlink channel at slot $t$ and subcarrier $n$ is represented as an $N_p$-dimensional vector and is denoted by
\begin{equation}
\mathbf{h}^{(f)}_{b,t}[n]\in\mathbb{C}^{N_p\times 1}.
\label{eq:h_freq_vec_rev}
\end{equation}
For exposition, it can be written in the geometric form
\begin{align}
\mathbf{h}^{(f)}_{b,t}[n]
\!\!=\!\!\!
\sum_{\ell=1}^{L_{b,t}}\!\!
\alpha_{b,\ell}(t)
e^{-j2\pi f_n\tau_{b,\ell}(t)}\!
e^{j2\pi \nu_{b,\ell}(t)t_s}
\mathbf{a}_b\!\left(\phi_{b,\ell}(t),\!\theta_{b,\ell}(t)\right),
\label{eq:geom_channel_rev}
\end{align}
where $L_{b,t}$ is the number of effective paths, $f_n$ denotes the frequency of the $n$-th active subcarrier, and $\alpha_{b,\ell}(t)$, $\tau_{b,\ell}(t)$, $\nu_{b,\ell}(t)$, $\phi_{b,\ell}(t)$, and $\theta_{b,\ell}(t)$ denote the complex gain, delay, Doppler shift, azimuth angle, and elevation angle of the $\ell$-th path, respectively. Under a slot-wise quasi-static approximation, these path parameters are treated as constant within slot $t$. This approximation is consistent with the slot-based transmission structure of 5G NR \cite{3gpp38211}, and is standard for slot-level beam prediction because the decision is made at the slot level rather than at the symbol level. It allows the model to focus on inter-slot evolution of the propagation state, which is exactly the timescale relevant to next-step beam tracking.

Let $N_f$ denote the number of active OFDM subcarriers used for CSI observation. Stacking the channel vectors over these $N_f$ subcarriers gives the wideband CSI matrix
\begin{equation}
\mathbf{H}^{(f)}_{b,t}
=
\left[
\mathbf{h}^{(f)}_{b,t}[1],\mathbf{h}^{(f)}_{b,t}[2],\ldots,\mathbf{h}^{(f)}_{b,t}[N_f]
\right]
\in\mathbb{C}^{N_p\times N_f}.
\label{eq:H_freq_mat_rev}
\end{equation}
In the considered time-division duplex (TDD) setting, the CSI collected from pilots is used as a proxy for the downlink-equivalent channel under reciprocity, so the predictor can operate continuously on historical multi-BS observations while targeting a future joint BS--beam label.

\subsection{Dual-View Cooperative CSI Representation}

Besides the frequency-domain CSI, we also use its delay-domain representation as an alternative structured view of the same wideband channel. This transformation does not create an independent measurement, since it is an invertible DFT-domain conversion. Its purpose is to present the same CSI in a form where delay-related sparsity and multipath concentration can be more easily captured by the subsequent CNN-based tokenizer. We therefore define
\begin{equation}
\mathbf{H}^{(\tau)}_{b,t}
=
\mathbf{H}^{(f)}_{b,t}\mathbf{F}_{N_f}^{H},
\label{eq:H_delay_rev}
\end{equation}
where $\mathbf{F}_{N_f}$ is the normalized $N_f$-point discrete Fourier transform (DFT) matrix, and equivalently
\begin{equation}
\mathbf{H}^{(f)}_{b,t}
=
\mathbf{H}^{(\tau)}_{b,t}\mathbf{F}_{N_f}.
\label{eq:H_freq_recover_rev}
\end{equation}

In our implementation, the dataset stores the delay-domain tensor and reconstructs the frequency-domain view by FFT during loading. Therefore, the two views should be understood as two equivalent domain representations of the same channel observation, rather than two independent sources of information.

For each BS and slot, we define the two-view observation as
\begin{equation}
\mathcal{H}_{b,t}
=
\left\{
\mathbf{H}^{(\tau)}_{b,t},\; \mathbf{H}^{(f)}_{b,t}
\right\}.
\label{eq:two_view_obs_rev}
\end{equation}
Let $T_h$ denote the history length. The cooperative observation window ending at slot $t$ is
\begin{equation}
\mathcal{X}_t
=
\left\{
\mathcal{H}_{b,\ell}
\right\}_{\substack{b=1,\ldots,N_{\mathrm{BS}}\\ \ell=t-T_h+1,\ldots,t}}.
\label{eq:coop_history_set_rev}
\end{equation}
Since the neural predictor operates on real-valued tensors, the complex CSI is separated into real and imaginary parts. The final model input is therefore organized as
\begin{equation}
\mathbf{X}_t
\in
\mathbb{R}^{T_h \times N_{\mathrm{BS}} \times N_{\mathrm{view}} \times 2 \times N_p \times N_f},
\label{eq:input_tensor_rev}
\end{equation}
where $N_{\mathrm{view}}=2$, and the dimensions correspond to time, BS index, CSI view, real/imaginary channels, antenna ports, and subcarriers, respectively.

\subsection{Beam Codebook and Joint BS--Beam Label Space}

Each BS uses a predefined DFT-based beam codebook with $N_{\mathrm{beam}}$ candidate beams, where $N_{\mathrm{beam}}$ denotes the number of beamforming vectors available at each BS. Let
\begin{equation}
\mathbf{C}_{b}
=
\left[
\mathbf{f}_{b,1},\mathbf{f}_{b,2},\ldots,\mathbf{f}_{b,N_{\mathrm{beam}}}
\right]
\in \mathbb{C}^{N_p\times N_{\mathrm{beam}}},
\label{eq:beam_codebook_matrix_rev}
\end{equation}
where the $m$-th candidate beam is given by
\begin{equation}
\mathbf{f}_{b,m}
=
\frac{1}{\sqrt{N_p}}
\left[
1,
e^{-j\frac{2\pi(m-1)}{N_{\mathrm{beam}}}},
\ldots,
e^{-j\frac{2\pi(N_p-1)(m-1)}{N_{\mathrm{beam}}}}
\right]^T .
\label{eq:dft_beam_vector_rev}
\end{equation}
The corresponding beam codebook is
\begin{equation}
\mathcal{F}_b
=
\left\{
\mathbf{f}_{b,1},\mathbf{f}_{b,2},\ldots,\mathbf{f}_{b,N_{\mathrm{beam}}}
\right\},
\label{eq:beam_codebook_rev}
\end{equation}
where $\mathbf{f}_{b,m}\in\mathbb{C}^{N_p\times 1}$ and $\|\mathbf{f}_{b,m}\|_2=1$.

For a BS--beam pair $(b,m)$, the wideband beam gain at slot $t$ is defined as
\begin{equation}
g_{b,m}(t)
=
\sum_{n=1}^{N_f}
\left|
\mathbf{f}_{b,m}^{H}\mathbf{h}^{(f)}_{b,t}[n]
\right|^2.
\label{eq:beam_gain_rev}
\end{equation}
The optimal pair is then
\begin{equation}
(b_t^{\star},m_t^{\star})
=
\arg\max_{\substack{1\le b\le N_{\mathrm{BS}}\\1\le m\le N_{\mathrm{beam}}}}
g_{b,m}(t).
\label{eq:best_pair_rev}
\end{equation}
Instead of using a hierarchical BS-then-beam predictor, we define a flat global label
\begin{equation}
y_t
=
(b_t^{\star}-1)N_{\mathrm{beam}} + m_t^{\star},
\quad
 y_t\in\{1,2,\ldots,C\},
\label{eq:joint_label_rev}
\end{equation}
where
\begin{equation}
C = N_{\mathrm{BS}}N_{\mathrm{beam}}
\label{eq:num_joint_classes_rev}
\end{equation}
is the total number of joint BS--beam classes. This flat formulation allows the model to handle both intra-BS beam switching and inter-BS handover in one prediction space.

\subsection{Received-Signal Interpretation}

The beam-gain label definition in \eqref{eq:beam_gain_rev} is consistent with the received signal model. If beam $\mathbf{f}_{b,m}$ is selected at slot $t$ and subcarrier $n$, the received signal is
\begin{equation}
r_{b,m,t}[n]
=
\mathbf{h}^{(f)H}_{b,t}[n]\mathbf{f}_{b,m}x_t[n] + z_t[n],
\label{eq:rx_signal_rev}
\end{equation}
where $x_t[n]$ is the transmitted symbol and $z_t[n]\sim\mathcal{CN}(0,\sigma^2)$ is additive noise, assumed to be independent of $x_t[n]$. Assuming
\begin{equation}
\mathbb{E}\!\left[|x_t[n]|^2\right]=P_x,
\label{eq:tx_symbol_power_rev}
\end{equation}
the corresponding average received power is
\begin{equation}
\mathbb{E}\!\left[|r_{b,m,t}[n]|^2\right]
=
\left|
\mathbf{h}^{(f)H}_{b,t}[n]\mathbf{f}_{b,m}
\right|^2 P_x + \sigma^2.
\label{eq:avg_rx_power_rev}
\end{equation}
Therefore, when the symbol power is identical across candidate beams and subcarriers, maximizing the received signal power is equivalent to maximizing $\left|\mathbf{h}^{(f)H}_{b,t}[n]\mathbf{f}_{b,m}\right|^2$, and the corresponding wideband label can be constructed using the summed gain in \eqref{eq:beam_gain_rev}.

\section{Problem Formulation}

\subsection{Cooperative Next-Step Joint BS--beam Prediction}

Given the historical cooperative CSI tensor $\mathbf{X}_t$ and the current joint label $y_t$, our goal is to predict the optimal joint BS--beam label at slot $t+\Delta$, where $\Delta\ge 1$ denotes the prediction horizon. In the main setting of this paper, we use $\Delta=1$. Here, $y_t$ is assumed to be available at inference time as the current confirmed joint BS--beam label obtained from the beam-management decision at slot $t$.

Let $\Phi_{\Theta}(\cdot)$ denote a trainable predictor parameterized by $\Theta$. The prediction mapping is written as
\begin{equation}
\mathbf{p}_{t+\Delta}
=
\Phi_{\Theta}(\mathbf{X}_t,y_t)
\in [0,1]^C,
\qquad
\sum_{c=1}^{C}\mathbf{p}_{t+\Delta}[c]=1,
\label{eq:predictor_mapping_rev}
\end{equation}
where $\mathbf{p}_{t+\Delta}$ is the predicted probability distribution over the $C$ joint classes. The hard prediction is
\begin{equation}
\hat{y}_{t+\Delta}
=
\arg\max_{1\le c\le C}\mathbf{p}_{t+\Delta}[c].
\label{eq:hard_prediction_rev}
\end{equation}
This formulation explicitly couples cooperative multi-BS observation and unified BS--beam decision making.

\subsection{Compact Candidate-Set Prediction}

In practical beam management, a compact set of strong candidates is often more useful than a single rigid decision. We therefore also consider the top-$K$ candidate set
\begin{equation}
\hat{\mathcal{S}}_{t+\Delta}^{(K)}
=
\operatorname{TopK}\!\left(\mathbf{p}_{t+\Delta},K\right),
\quad K\ll C,
\label{eq:topk_candidate_set_rev}
\end{equation}
where $\operatorname{TopK}(\cdot)$ returns the indices of the $K$ largest entries in $\mathbf{p}_{t+\Delta}$. This perspective is especially important in high-mobility and blockage-sensitive scenarios, where several nearby beams or multiple BSs may all be good candidates at the same time. In practice, such a candidate-set view is also compatible with low-overhead beam refinement procedures, because the system only needs to verify a small number of strong alternatives rather than re-scan the entire joint search space. Therefore, Top-$K$ prediction is not merely an auxiliary metric, but also a practically meaningful operating mode for cooperative beam management.

\subsection{Stable and Abrupt-Change Beam Evolution}

\begin{figure}[t]
    \centering
    \includegraphics[width=\linewidth]{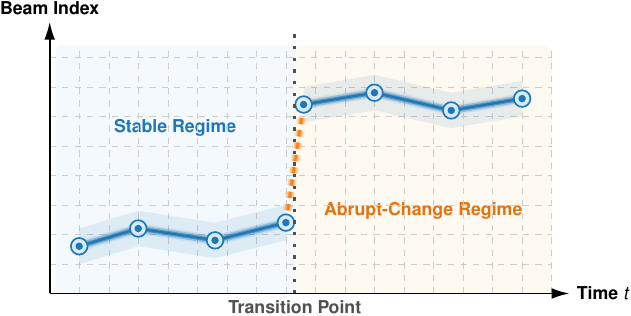}
    \caption{Stable and abrupt-change beam evolution. The optimal beam may evolve smoothly for several slots and then change sharply near a transition point.}
    \label{fig:motivation}
\end{figure}

As illustrated in Fig.~\ref{fig:motivation}, the optimal beam sequence may remain unchanged for several slots and then switch abruptly near a transition point. To characterize this behavior, we define the transition indicator
\begin{equation}
s_{t+\Delta}
=
\mathbb{I}\!\left\{ y_{t+\Delta}\neq y_t \right\},
\label{eq:transition_indicator_rev}
\end{equation}
where $s_{t+\Delta}=0$ corresponds to the stable regime and $s_{t+\Delta}=1$ corresponds to the abrupt-change regime.

This regime distinction motivates the output design of the predictor: stable samples mainly require continuation of the current beam trajectory, whereas transition samples require sensitivity to sharper changes induced by blockage, road turning, or BS handover.

\subsection{Training Dataset and Learning Objective}

Let
\begin{equation}
\mathcal{D}
=
\left\{
\big(\mathbf{X}_t^{(i)}, y_t^{(i)}, y_{t+\Delta}^{(i)}\big)
\right\}_{i=1}^{N_{\mathrm{tr}}}
\label{eq:training_dataset_rev}
\end{equation}
denote the training set, where $\mathbf{X}_t^{(i)}$ is the historical cooperative CSI tensor, $y_t^{(i)}$ is the current joint label, and $y_{t+\Delta}^{(i)}$ is the ground-truth future label. The standard supervised objective is the cross-entropy loss
\begin{equation}
\mathcal{L}_{\mathrm{CE}}
=
-\frac{1}{N_{\mathrm{tr}}}
\sum_{i=1}^{N_{\mathrm{tr}}}
\log \mathbf{p}_{t+\Delta}^{(i)}\!\left[y_{t+\Delta}^{(i)}\right],
\label{eq:ce_loss_rev}
\end{equation}
or equivalently,
\begin{equation}
\mathcal{L}_{\mathrm{CE}}
=
-\frac{1}{N_{\mathrm{tr}}}
\sum_{i=1}^{N_{\mathrm{tr}}}
\sum_{c=1}^{C}
\mathbf{q}_{t+\Delta}^{(i)}[c]\log \mathbf{p}_{t+\Delta}^{(i)}[c],
\label{eq:ce_loss_onehot_rev}
\end{equation}
where $\mathbf{q}_{t+\Delta}^{(i)}$ is the one-hot label vector. The learning problem is thus
\begin{equation}
\min_{\Theta}\; \mathcal{L}_{\mathrm{CE}}.
\label{eq:learning_problem_rev}
\end{equation}

\section{Proposed CRS-LLM Architecture}

In this section, we present the proposed CRS-LLM architecture for next-step joint BS--beam prediction. The model converts the high-dimensional historical cooperative dual-view CSI tensor into a compact token sequence, models temporal evolution with a truncated GPT-style backbone, and produces the next-step beam distribution through a transition-aware switch-gated predictor. Compared with conventional sequence classifiers, the proposed framework is designed for three main characteristics of this task: cooperative multi-BS observation, dual-view CSI representation, and the coexistence of smooth evolution and abrupt beam transitions.

\subsection{Framework Overview}

\begin{figure*}[t]
    \centering
    \includegraphics[width=\textwidth]{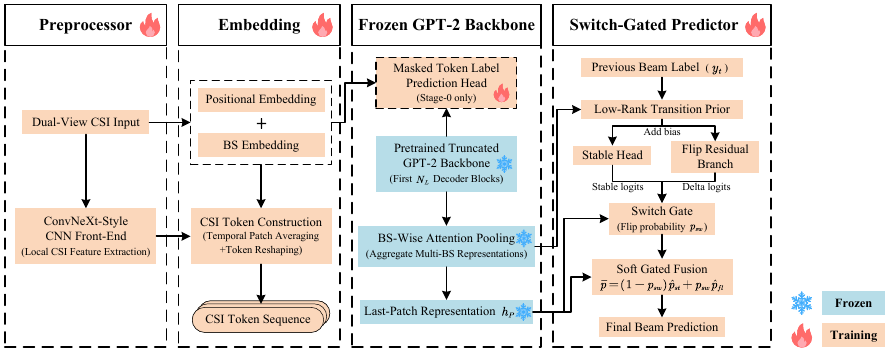}
    \caption{Overall architecture of the proposed CRS-LLM framework. The model consists of a preprocessor, an embedding module, a truncated GPT-style backbone, and a switch-gated predictor.}
    \label{fig:crsllm_overall}
\end{figure*}

The overall architecture is shown in Fig.~\ref{fig:crsllm_overall}. Starting from the cooperative CSI tensor in \eqref{eq:input_tensor_rev}, the model first applies a lightweight CNN front-end to each BS--view--time slice in order to compress the local spatial--frequency structure of CSI. The resulting features are then projected into a token space and organized along the temporal axis. A truncated GPT-style backbone models the cross-time dependence of these tokens, after which a BS-wise attention pooling module aggregates multi-BS information patch by patch. Finally, a switch-gated predictor produces the next-step joint BS--beam distribution by combining a stable branch, a residual flip branch, and a transition prior induced by the previous beam label.

The framework is modular. The front-end performs CSI compression, the transformer backbone models temporal dependence, and the prediction head captures continuation and transition patterns. Such a decomposition is useful because the three subproblems have different inductive biases: local CSI compression is mainly spatial--frequency in nature, temporal evolution requires sequence modeling, and final beam decision making must remain sensitive to structured transitions in the joint BS--beam space. Separating these roles makes the model easier to interpret and avoids asking one module to learn everything at once.

\subsection{Cooperative CSI Tokenization and Embedding}

Recall that the model input is
\begin{equation}
\mathbf{X}_t
\in
\mathbb{R}^{T_h \times N_{\mathrm{BS}} \times N_{\mathrm{view}} \times 2 \times N_p \times N_f},
\label{eq:method_input_tensor_rev}
\end{equation}
where the dimensions correspond to the history length, BS identity, CSI view, real/imaginary channels, antenna ports, and active OFDM subcarriers.

\subsubsection{Local CSI front-end}

The raw CSI tensor is high-dimensional and exhibits strong local correlation over the antenna-port and subcarrier dimensions. Instead of feeding the raw tensor directly into the transformer, we first process each BS--view--time slice using a lightweight ConvNeXt-style CNN front-end. For BS $b$, view $v$, and time index $\tau$, let
\begin{equation}
\mathbf{X}_{\tau,b,v}
=
\mathbf{X}_t[\tau,b,v,:,:,:]
\in
\mathbb{R}^{2\times N_p \times N_f}.
\label{eq:xbv_tau_rev}
\end{equation}
Before being fed into the CNN front-end, each CSI slice is normalized by instance normalization over its antenna-port and subcarrier dimensions. This reduces the scale variation among different BSs, views, and time slots while preserving the local spatial--frequency pattern. The front-end then computes
\begin{equation}
\mathbf{u}_{\tau,b,v}
=
\Psi_{\mathrm{cnn}}(\mathbf{X}_{\tau,b,v})
\in
\mathbb{R}^{d_c},
\label{eq:cnn_frontend_rev}
\end{equation}
where $d_c$ is the compact feature dimension. This step suppresses local redundancy and converts each CSI slice into a dense representation that is easier for the transformer to model.

\subsubsection{Temporal patching and feature projection}

To organize the historical CSI features into a transformer-compatible token sequence, the history window is divided into temporal patches of length $L_p$. When $L_p>1$, this operation also reduces the sequence length by aggregating neighboring time steps. In the experiments of this paper, we set $L_p=1$, so no temporal downsampling is applied and each slot corresponds to one temporal patch. The number of patches is
\begin{equation}
N_{\mathrm{patch}}
=
\frac{T_h}{L_p},
\label{eq:num_patch_rev}
\end{equation}
where $T_h$ is divisible by $L_p$ in our implementation. For the $p$-th patch, we compute the patch feature by averaging the front-end outputs within the patch:
\begin{equation}
\bar{\mathbf{u}}_{b,v,p}
=
\frac{1}{L_p}
\sum_{\tau\in\mathcal{T}_p}
\mathbf{u}_{\tau,b,v}
\in \mathbb{R}^{d_c},
\label{eq:patch_average_rev}
\end{equation}
where $\mathcal{T}_p$ is the index set of the $p$-th temporal patch. For $L_p=1$, this averaging operation reduces to using the original slot-level feature directly. These patch features are then projected into the transformer embedding space:
\begin{equation}
\tilde{\mathbf{e}}_{b,v,p}
=
\mathbf{W}_{e}\bar{\mathbf{u}}_{b,v,p}+\mathbf{b}_{e}
\in \mathbb{R}^{d}.
\label{eq:linear_projection_rev}
\end{equation}
The patch-based formulation is retained because it also covers straightforward extensions to longer temporal aggregation windows.

\subsubsection{Dual-view representation and structural embedding}

The delay-domain and frequency-domain inputs are two equivalent views of the same wideband CSI. Their projected embeddings are fused by summation before entering the transformer:
\begin{equation}
\tilde{\mathbf{e}}_{b,p}
=
\tilde{\mathbf{e}}_{b,d,p}+\tilde{\mathbf{e}}_{b,f,p},
\label{eq:view_sum_fusion_rev}
\end{equation}
where the subscripts $d$ and $f$ denote the delay-domain and frequency-domain representations, respectively. We use additive fusion because it keeps the token dimension unchanged and encourages the two views to remain in a shared feature space before temporal modeling.

To encode structural information, we add a temporal positional embedding and a BS-specific embedding:
\begin{equation}
\mathbf{e}_{b,p}
=
\tilde{\mathbf{e}}_{b,p}
+
\mathbf{e}^{\mathrm{pos}}_{p}
+
\mathbf{e}^{\mathrm{BS}}_{b}.
\label{eq:compound_embedding_rev}
\end{equation}
For the default summed-view configuration, all tokens are concatenated into the cooperative sequence
\begin{align}
\mathbf{E}_t
=&
\left[
\mathbf{e}_{1,1},\ldots,\mathbf{e}_{N_{\mathrm{BS}},1},\ldots,\mathbf{e}_{1,N_{\mathrm{patch}}},\ldots,\mathbf{e}_{N_{\mathrm{BS}},N_{\mathrm{patch}}}
\right]\nonumber\\
&\in\mathbb{R}^{N_{\mathrm{tok}}\times d},
\label{eq:token_sequence_rev}
\end{align}
with total sequence length $N_{\mathrm{tok}} = N_{\mathrm{BS}}N_{\mathrm{patch}}$.
This tokenization step converts raw CSI into a sequence whose ordering and identity reflect the physical structure of the beam tracking problem, making it more suitable for transformer-based temporal modeling.

\subsection{Truncated GPT-Style Temporal Backbone}

The cooperative token sequence $\mathbf{E}_t$ is processed by a truncated GPT-2 backbone consisting of the first $N_L$ decoder blocks of a pre-trained model. Denoting this backbone by $\mathrm{GPT}_{1:N_L}(\cdot)$, the contextualized hidden sequence is
\begin{equation}
\mathbf{H}_t
=
\mathrm{GPT}_{1:N_L}(\mathbf{E}_t)
\in\mathbb{R}^{N_{\mathrm{tok}}\times d}.
\label{eq:gpt_hidden_rev}
\end{equation}

The original GPT positional embedding table is disabled, and the backbone is frozen except for normalization-related parameters. The specific value of $N_L$ is treated as an experimental hyperparameter and is reported in the experimental settings. This parameter-efficient configuration preserves the temporal prior of the pre-trained backbone while reducing the risk of overfitting on the beam prediction dataset. The CSI front-end, structural embeddings, and switch-gated head remain trainable, which provides sufficient flexibility for adaptation to the wireless task and contributes to stable optimization.

\subsubsection{Patch-wise BS attention pooling}

After the transformer, the hidden sequence is reshaped back to a patch-by-BS structure. Let $\mathbf{h}_{b,p}\in\mathbb{R}^{d}$ denote the contextualized token corresponding to BS $b$ and patch $p$. Since the final target is a single joint BS--beam label, the multi-BS features need to be summarized into a compact cooperative representation. Instead of using simple averaging, we adopt attention pooling over the BS dimension. This allows the model to assign larger weights to more informative BSs while still retaining information from other BSs:
\begin{align}
\alpha_{b,p}
&=
\frac{
\exp\!\left(\mathbf{a}^{T}\tanh(\mathbf{W}_{a}\mathbf{h}_{b,p})\right)
}{
\sum\limits_{j=1}^{N_{\mathrm{BS}}}
\exp\!\left(\mathbf{a}^{T}\tanh(\mathbf{W}_{a}\mathbf{h}_{j,p})\right)
},
\label{eq:bs_attention_weight_rev}
\\
\mathbf{r}_{p}
&=
\sum_{b=1}^{N_{\mathrm{BS}}}
\alpha_{b,p}\mathbf{h}_{b,p}.
\label{eq:patch_representation_rev}
\end{align}
Compared with uniform averaging, this attention-based aggregation reduces the multi-BS feature dimension while keeping the BS contribution adaptive to the current channel state. This yields a patch-wise representation sequence $\{\mathbf{r}_1,\ldots,\mathbf{r}_{N_{\mathrm{patch}}}\}$. 
We use the last patch representation
\begin{equation}
\mathbf{h}_{p}=\mathbf{r}_{N_{\mathrm{patch}}}
\label{eq:last_step_representation_rev}
\end{equation}
as the global representation for next-step prediction. This choice is motivated by the autoregressive nature of the task: the next-step decision is most directly related to the latest observed temporal context, while earlier patches have already been integrated into the hidden dynamics of the GPT backbone. Compared with uniform averaging over all patches, using the final patch representation places greater emphasis on the most recent cooperative CSI state without discarding the preceding temporal information.

\subsection{Transition-Aware Switch-Gated Predictor}

\begin{figure*}[t]
    \centering
    \includegraphics[width=0.8\textwidth]{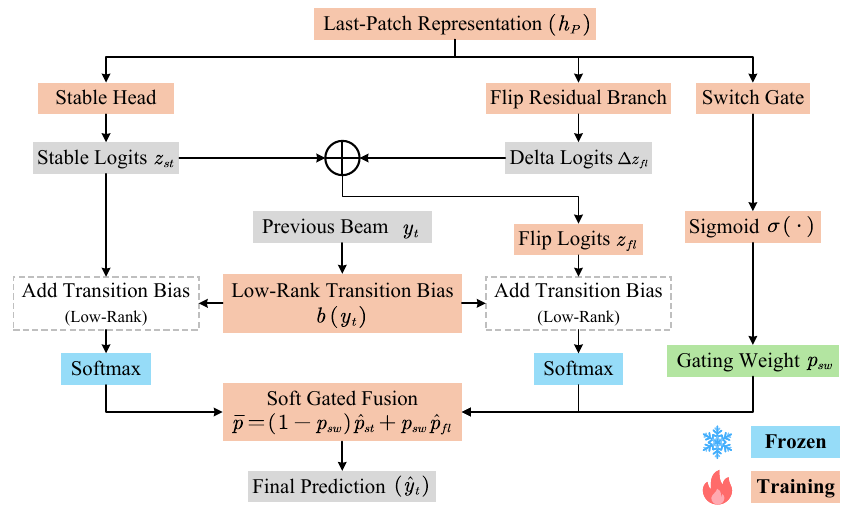}
    \caption{Detailed structure of the switch-gated predictor. The predictor combines a stable head, a flip residual branch, a low-rank transition bias from the previous beam label, and a soft gated fusion mechanism.}
    \label{fig:switch_gated_predictor}
\end{figure*}

The output head is illustrated in Fig.~\ref{fig:switch_gated_predictor}. Instead of using a single classification head, we explicitly decompose the prediction into a continuation-oriented component and a transition-oriented correction component. This design is motivated by the physical characteristics of beam evolution in V2X scenarios. Stable samples usually follow local trajectory continuity and beam persistence, whereas transition samples are more strongly affected by blockage, road turning, or inter-BS handover. These two regimes need not share the same decision pattern, and forcing them into a single prediction head may weaken sensitivity to rare but important transitions.

\subsubsection{Stable head and residual flip branch}

The stable branch produces the base logits
\begin{equation}
\mathbf{z}_{\mathrm{st}}
=
\mathbf{W}_{\mathrm{st}}\mathbf{h}_{p}+\mathbf{b}_{\mathrm{st}}
\in\mathbb{R}^{C}.
\label{eq:stable_logits_rev}
\end{equation}
To model abrupt changes, we add a residual flip branch
\begin{equation}
\Delta \mathbf{z}_{\mathrm{fl}}
=
\mathbf{W}_{\Delta}\mathbf{h}_{p}+\mathbf{b}_{\Delta}
\in\mathbb{R}^{C},
\label{eq:flip_delta_rev}
\end{equation}
and define
\begin{equation}
\mathbf{z}_{\mathrm{fl}}
=
\mathbf{z}_{\mathrm{st}} + \Delta \mathbf{z}_{\mathrm{fl}}.
\label{eq:flip_logits_rev}
\end{equation}
This residual design allows the flip branch to correct the stable prediction instead of learning a completely new class distribution from scratch. It is therefore more parameter-efficient and is consistent with the observation that many transition samples can be interpreted as local corrections to an otherwise stable beam trajectory.

\subsubsection{Low-rank transition prior from the previous label}

Let $\mathbf{q}_t\in\{0,1\}^{C}$ be the one-hot encoding of the current label $y_t$. We construct a low-rank transition prior
\begin{equation}
\mathbf{b}(y_t)
=
\mathbf{U}\mathbf{V}^{T}\mathbf{q}_t
\in\mathbb{R}^{C},
\label{eq:low_rank_transition_prior_rev}
\end{equation}
where $\mathbf{U},\mathbf{V}\in\mathbb{R}^{C\times r}$ and $r\ll C$. This prior can be interpreted as a low-rank interaction between the previous class and all candidate classes. The prior is added to both branches:
\begin{align}
\tilde{\mathbf{z}}_{\mathrm{st}}
&=
\mathbf{z}_{\mathrm{st}} + \mathbf{b}(y_t),
\label{eq:stable_bias_added_rev}
\\
\tilde{\mathbf{z}}_{\mathrm{fl}}
&=
\mathbf{z}_{\mathrm{fl}} + \mathbf{b}(y_t).
\label{eq:flip_bias_added_rev}
\end{align}
The corresponding branch distributions are
\begin{align}
\hat{\mathbf{p}}_{\mathrm{st}}
&=
\mathrm{softmax}(\tilde{\mathbf{z}}_{\mathrm{st}}),
\label{eq:p_st_rev}
\\
\hat{\mathbf{p}}_{\mathrm{fl}}
&=
\mathrm{softmax}(\tilde{\mathbf{z}}_{\mathrm{fl}}).
\label{eq:p_fl_rev}
\end{align}
This prior provides a compact way to encode how the next label depends on the current label in the joint BS--beam space. In other words, once the current label is known, the future label distribution is typically highly structured rather than uniform: some candidate BS--beam pairs are likely neighboring candidates or handover targets, whereas many others are unlikely to occur in the next step.

\subsubsection{Switch gate and soft fusion}

A switch gate estimates the probability that the next-step evolution belongs to the abrupt-change regime:
\begin{equation}
p_{\mathrm{sw}}
=
\sigma\!\left(\mathbf{w}_{g}^{T}\mathbf{h}_{p}+b_g\right)
\in(0,1),
\label{eq:switch_probability_rev}
\end{equation}
where $\sigma(\cdot)$ is the sigmoid function. The final prediction is obtained by probability-level soft fusion:
\begin{equation}
\bar{\mathbf{p}}_{t+\Delta}
=
(1-p_{\mathrm{sw}})\hat{\mathbf{p}}_{\mathrm{st}}
+
p_{\mathrm{sw}}\hat{\mathbf{p}}_{\mathrm{fl}}.
\label{eq:soft_gated_fusion_rev}
\end{equation}
The final class prediction is
\begin{equation}
\hat{y}_{t+\Delta}
=
\arg\max_{1\le c\le C}
\bar{\mathbf{p}}_{t+\Delta}[c].
\label{eq:final_prediction_method_rev}
\end{equation}
Soft fusion allows the model to move smoothly between continuation-oriented and transition-oriented predictions, which makes it less sensitive to unclear boundaries between the two regimes.

\subsection{Training Strategy}

The model is trained in two stages in the main pipeline.

\subsubsection{Stage 0: masked historical-label warm-up}

Before next-step supervision, we warm up the trainable modules with a masked temporal-token task. Let $\mathbf{Y}^{\mathrm{hist}}\in\{1,\ldots,C\}^{T_h}$ denote the historical joint-label sequence within the input window. A subset $\mathcal{M}$ of temporal positions is masked, and the corresponding CSI slices are perturbed with an 80/10/10 rule: 80\% are replaced by zeros, 10\% are replaced by randomly selected positions, and the remaining 10\% are kept unchanged. In the main configuration $L_p=1$, each masked temporal position corresponds to one temporal patch. After BS-wise aggregation, the patch-wise aggregated representation $\mathbf{r}_j$ of each masked patch $j\in\mathcal{M}$ is fed into an auxiliary token-level head to predict the corresponding historical joint label. If $\tilde{\mathbf{p}}_{j}$ denotes the predicted class distribution for the $j$-th masked time step, the stage-0 objective is
\begin{equation}
\mathcal{L}_{\mathrm{mask}}
=
-\frac{1}{|\mathcal{M}|}
\sum_{j\in\mathcal{M}}
\log \tilde{\mathbf{p}}_{j}[y_j],
\label{eq:masked_pretraining_loss_rev}
\end{equation}
where $y_j$ is the ground-truth historical joint label at the $j$-th masked time step. This stage does not directly optimize the final prediction objective. Instead, it provides a milder supervision signal that encourages the trainable tokenizer and aggregation layers to first learn beam-evolution-related structure from historical observations. In this sense, stage-0 warm-up serves as a task-adapted initialization step: it narrows the modality gap between cooperative CSI tokens and the frozen GPT backbone before the more difficult next-step prediction objective is introduced.

\subsubsection{Stage 1: next-step beam prediction with switch supervision}

After the masked warm-up, the auxiliary head is removed and the full model is optimized for next-step prediction. The main beam loss is
\begin{equation}
\mathcal{L}_{\mathrm{beam}}
=
-\log \bar{\mathbf{p}}_{t+\Delta}[y_{t+\Delta}].
\label{eq:beam_prediction_loss_rev}
\end{equation}
To supervise the gate, we use the transition indicator $s_{t+\Delta}$ defined in \eqref{eq:transition_indicator_rev}, and define the gate loss as
\begin{equation}
\mathcal{L}_{\mathrm{sw}}
=
-
s_{t+\Delta}\log p_{\mathrm{sw}}
-
(1-s_{t+\Delta})\log(1-p_{\mathrm{sw}}).
\label{eq:gate_bce_loss_rev}
\end{equation}
The final stage-1 objective is
\begin{equation}
\mathcal{L}
=
\mathcal{L}_{\mathrm{beam}} + \lambda_{\mathrm{sw}}\mathcal{L}_{\mathrm{sw}},
\label{eq:final_training_loss_rev}
\end{equation}
where $\lambda_{\mathrm{sw}}>0$ balances class prediction and transition discrimination.

During stage-1 training, the truncated GPT backbone remains frozen except for normalization-related parameters, whereas the CNN front-end, projection layers, structural embeddings, BS-attention pooling, transition-prior module, switch gate, and prediction heads are trainable. This parameter-efficient setting is consistent with the experimental protocol and contributes to stable optimization.

\section{Experiments}

\subsection{Experimental Settings}

This section evaluates CRS-LLM from several aspects, including in-domain prediction performance, beam-gain preservation, training behavior, architectural contribution, data efficiency, regime-wise behavior, and cross-scenario transferability. All methods are tested under the same cooperative next-step joint BS--beam prediction protocol introduced in Section II. The prediction target is always the future joint class in the flattened label space of size $C = N_{\mathrm{BS}} N_{\mathrm{beam}}$.

The training and test samples are generated by a cooperative multi-BS mmWave channel simulator based on QuaDRiGa. In the main urban micro (UMi) setting, the system uses $N_{\mathrm{BS}}=4$ BSs, a history length of $T_h=16$ slots, a prediction horizon of $\Delta=1$, $N_p=32$ effective BS ports, and $N_f=64$ active subcarriers. A DFT-based codebook with $N_{\mathrm{beam}}=32$ beams is used at each BS, which gives $C=128$ joint BS--beam classes. For CRS-LLM, the truncated GPT-style backbone keeps the first $N_L=6$ decoder blocks of GPT-2, and the original GPT positional embedding table is disabled. In addition to CSI and future labels, the dataset also stores the full gain vector over all joint classes. This allows direct computation of NBG and helps show whether a predicted candidate set remains physically useful even when the exact Top-1 class is not selected.

We compare CRS-LLM with six baselines: CSI-Transformer, Hierarchical BS--Beam, CNN, RNN, LSTM, and GRU. All baselines use the same cooperative CSI input tensor, the same joint-label target, and the same train/validation/test split. We also keep the optimization budget comparable across methods. Their main hyperparameters are tuned in reasonable ranges to avoid underestimating baseline performance. In this way, the comparison focuses on differences in model design rather than differences in training setup.

We report both Top-1 accuracy and NBG. Since practical beam management often uses a short candidate list instead of only one hard decision, we also report Top-$K$ accuracy and NBG-$K$ for $K\in\{1,2,3,5\}$ when needed. The in-domain experiments cover SNR values from $-10$ dB to $20$ dB. Cross-scenario robustness is further tested by directly evaluating the trained model in an unseen urban macro (UMa) environment without any additional fine-tuning.

This setup evaluates not only exact future-class prediction, but also beam-gain preservation, optimization behavior, low-data performance, and transfer to an unseen scenario. Together, these tests provide a more complete view of the practical value of CRS-LLM.

\subsection{Evaluation Metrics}

We evaluate performance using both exact classification accuracy and beam-gain-oriented metrics. For a test sample $i$, the Top-$K$ hit indicator is
\begin{equation}
\mathrm{Acc}_{K}^{(i)}
=
\mathbb{I}
\left\{
y_{t+\Delta}^{(i)}\in \hat{\mathcal{S}}_{t+\Delta}^{(K),(i)}
\right\},
\label{eq:acc_k_rev}
\end{equation}
and the average Top-$K$ accuracy is
\begin{equation}
\mathrm{Acc}_{K}
=
\frac{1}{N_{\mathrm{te}}}
\sum_{i=1}^{N_{\mathrm{te}}}
\mathrm{Acc}_{K}^{(i)}.
\label{eq:avg_acc_k_rev}
\end{equation}

To measure how much physical beamforming gain is preserved by the predicted candidate set, we use the normalized beam gain (NBG). Let $g_c^{(i)}(t+\Delta)$ be the true beam gain of class $c$ for sample $i$. We define
\begin{equation}
\mathrm{NBG}_{K}^{(i)}
=
\frac{
\max\limits_{c\in \hat{\mathcal{S}}_{t+\Delta}^{(K),(i)}} g_c^{(i)}(t+\Delta)
}{
\max\limits_{1\le c\le C} g_c^{(i)}(t+\Delta)
},
\label{eq:nbg_at_k_rev}
\end{equation}
and the average value
\begin{equation}
\mathrm{NBG}_{K}
=
\frac{1}{N_{\mathrm{te}}}
\sum_{i=1}^{N_{\mathrm{te}}}
\mathrm{NBG}_{K}^{(i)}.
\label{eq:avg_nbg_at_k_rev}
\end{equation}
In our dataset, the gain vector over all joint classes is precomputed and stored for each sample and slot, so the NBG evaluation is directly linked to the physical codebook gain.

Since the task contains both stable continuation and abrupt transitions, we also evaluate these two regimes separately:
\begin{align}
\mathcal{D}_{\mathrm{st}}
&=
\left\{
i:\; y_{t+\Delta}^{(i)} = y_t^{(i)}
\right\},
\label{eq:stable_subset_rev}
\\
\mathcal{D}_{\mathrm{fl}}
&=
\left\{
i:\; y_{t+\Delta}^{(i)} \neq y_t^{(i)}
\right\}.
\label{eq:flip_subset_rev}
\end{align}
The corresponding values of $\mathrm{Acc}_{K}$ and $\mathrm{NBG}_{K}$ are then computed on these two subsets. This regime-wise view helps show how the model behaves under continuation-dominated and transition-dominated conditions.

\subsection{Overall Performance under Different SNR Conditions}

\begin{figure*}[t]
    \centering
    \begin{minipage}[b]{0.48\textwidth}
        \centering
        \includegraphics[width=\linewidth]{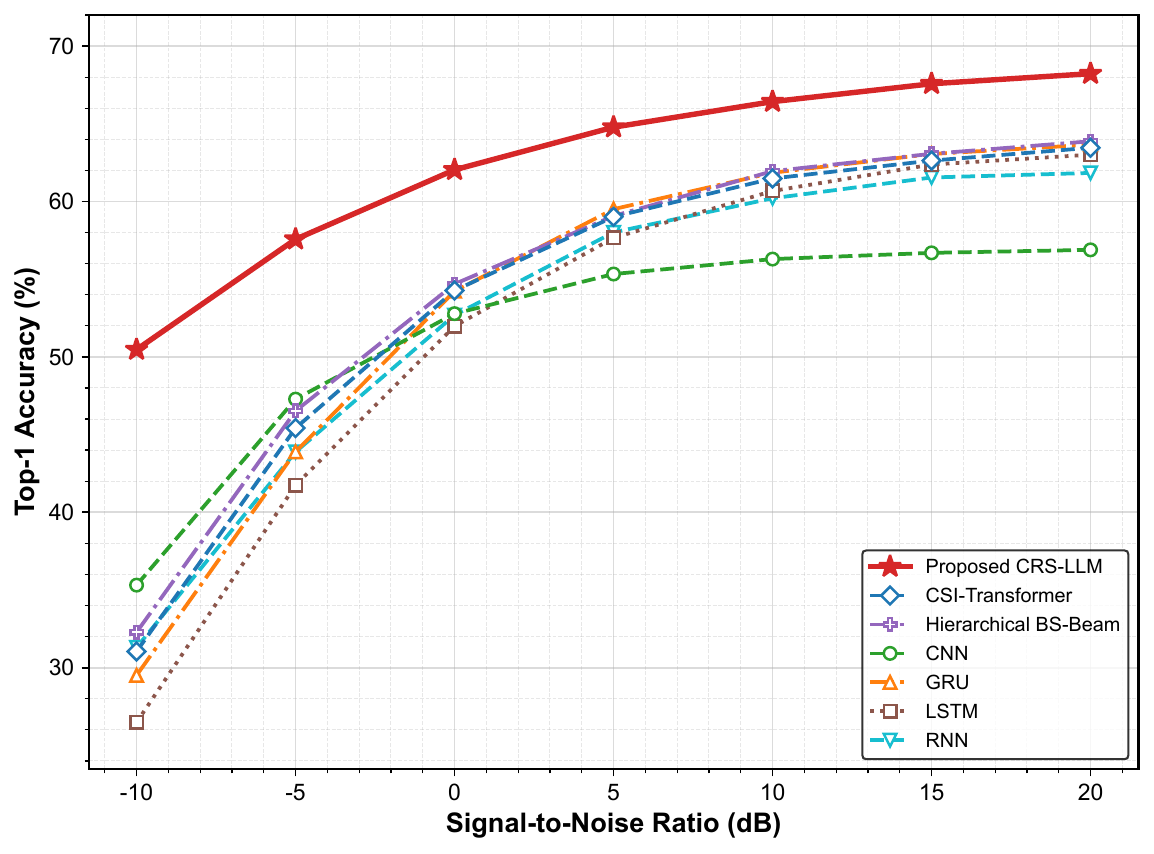}
    \end{minipage}
    \hfill
    \begin{minipage}[b]{0.48\textwidth}
        \centering
        \includegraphics[width=\linewidth]{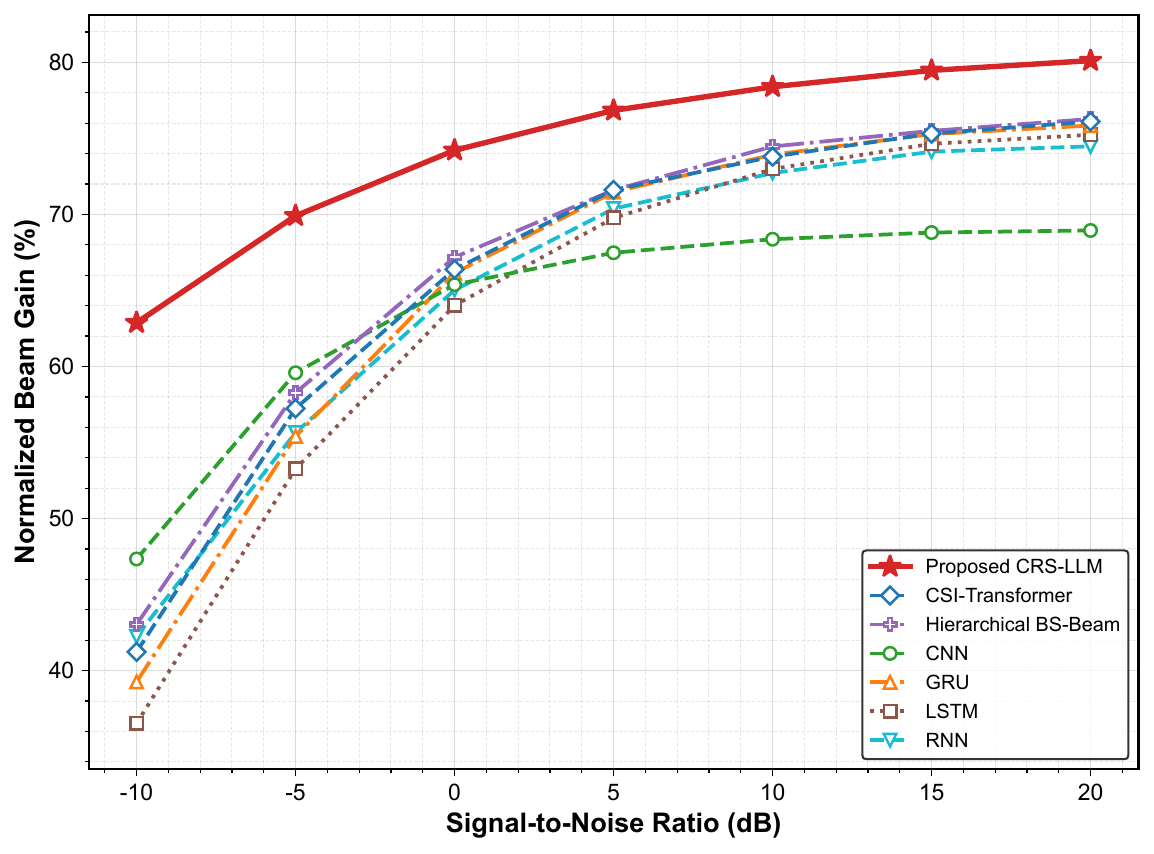}
    \end{minipage}
    \caption{Overall in-domain performance under different SNR conditions: left, Top-1 accuracy; right, NBG.}
    \label{fig:snr_main_natural}
\end{figure*}

We first evaluate the in-domain performance of CRS-LLM under different SNR conditions. The Top-1 accuracy and NBG results are shown in Fig.~\ref{fig:snr_main_natural}.

CRS-LLM achieves the best performance across the full SNR range, and the gain is especially clear at low SNR. At $-10$ dB, CRS-LLM still reaches about $50\%$ Top-1 accuracy, while the baselines stay around $27\%\sim35\%$. A similar gap appears in NBG. At the same SNR, CRS-LLM preserves about $63\%$ of the optimal beam gain, while the baselines remain below about $48\%$. This result is important because low-SNR prediction is not just a noisier version of the high-SNR case. When the CSI becomes less reliable, the model has to rely more on stable temporal and cross-BS structure instead of only local instantaneous details. The clear gap at low SNR therefore suggests that the proposed model can still extract useful information when the input becomes more difficult.

As SNR increases, all methods improve, but the relative ranking stays almost unchanged. CRS-LLM remains ahead of both CSI-Transformer and Hierarchical BS--Beam, as well as CNN-, RNN-, LSTM-, and GRU-based baselines. This suggests that the gain of CRS-LLM does not come only from better noise tolerance. It also comes from better modeling of temporal dependence and cooperative multi-BS information. In particular, the stronger results over CSI-Transformer suggest that the benefit does not come from using a transformer alone, while the gain over Hierarchical BS--Beam supports the value of direct joint prediction.

The NBG curves provide a practical view beyond exact classification accuracy. A Top-1 mismatch does not always mean a severe beam-selection error, because the predicted class may still remain close to the optimal one. The higher NBG of CRS-LLM therefore shows that its predicted ranking often remains useful in practice.

Overall, the SNR results show that CRS-LLM is more robust at low SNR and makes better use of structured temporal information at higher SNR.

\subsection{Convergence Behavior with and without Stage-0 Warm-Up}

\begin{figure}[h]
    \centering
    \includegraphics[width=\linewidth]{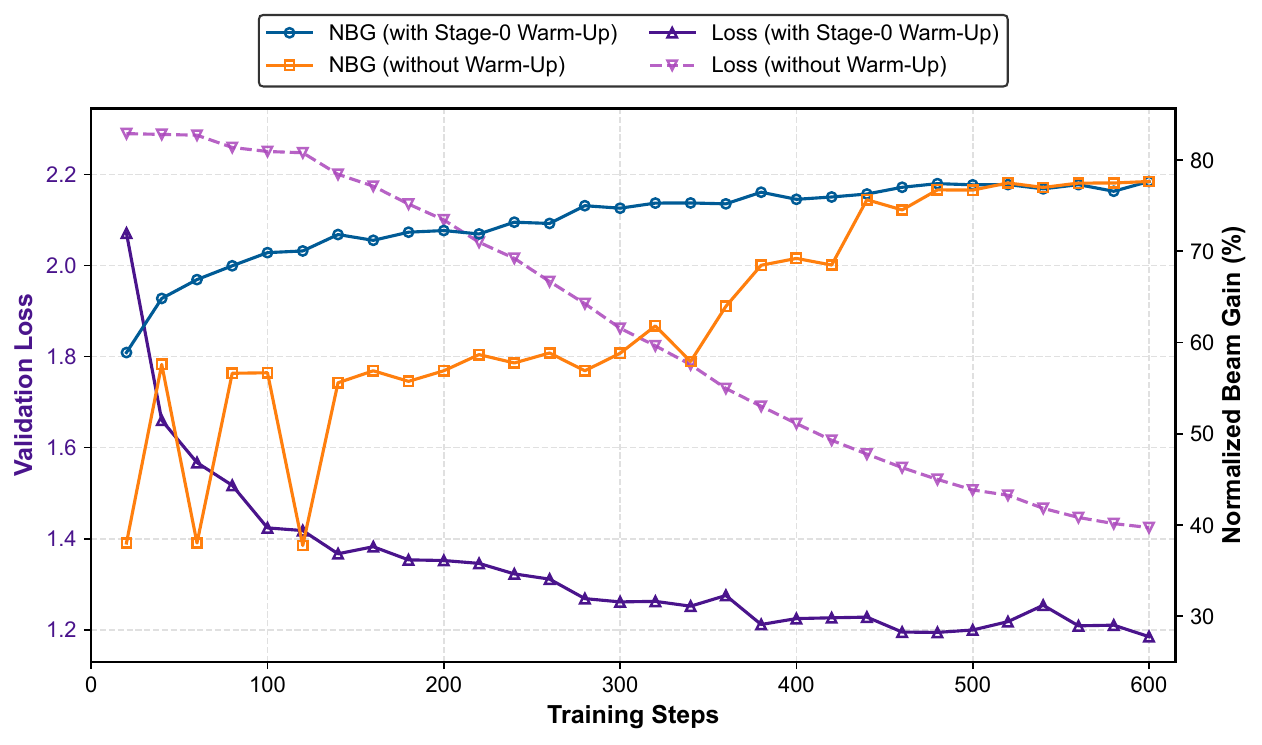}
    \caption{Convergence comparison between CRS-LLM with Stage-0 masked warm-up and CRS-LLM trained directly on the downstream next-step prediction task, in terms of validation loss and NBG.}
    \label{fig:convergence_natural}
\end{figure}

We next examine the optimization behavior of CRS-LLM. A key design choice of the training pipeline is to introduce a stage-0 masked warm-up before the final next-step prediction training. To evaluate its effect, Fig.~\ref{fig:convergence_natural} compares two settings under the same model architecture: 1) stage-0 warm-up followed by stage-1 next-step training, and 2) direct stage-1 training without the warm-up stage. The corresponding validation-loss and NBG curves are shown in Fig.~\ref{fig:convergence_natural}.

The model with warm-up shows a smoother and more stable training process than the model trained directly on the downstream task. Compared with direct training, it starts from a better operating point, improves more steadily, and converges to higher NBG with lower validation loss. This suggests that the stage-0 objective helps the trainable CSI tokenizer, structural embeddings, and aggregation layers learn beam-evolution-related structure before the harder next-step prediction objective is introduced.

This effect is especially relevant because cooperative CSI is high-dimensional and very different from natural-language tokens. The masked warm-up stage provides an easier intermediate signal, which helps align CSI features with the frozen GPT-style backbone, reduces optimization difficulty, and leads to more reliable stage-1 training.

Overall, the stage-0 warm-up serves as a useful initialization step that improves stability and downstream prediction performance.

\subsection{Ablation Study of the Switch-Gated Predictor}

\begin{figure}[h]
    \centering
    \includegraphics[width=\linewidth]{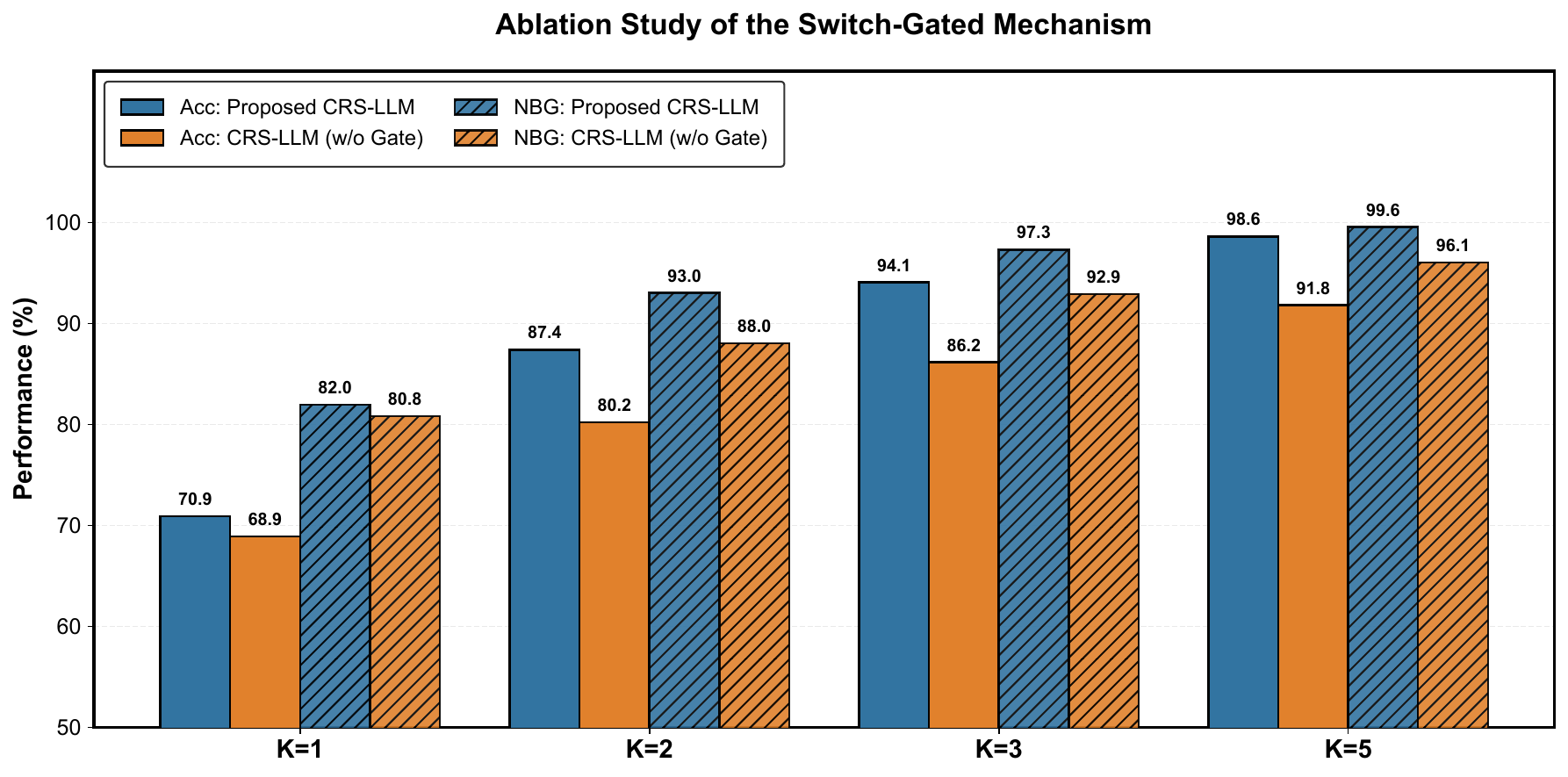}
    \caption{Ablation study of the switch-gated mechanism in terms of Top-$K$ accuracy and NBG.}
    \label{fig:ablation_gate_natural}
\end{figure}

We next isolate the contribution of the switch-gated output design by comparing the full CRS-LLM with a simplified ungated variant. The corresponding Top-$K$ accuracy and NBG results are shown in Fig.~\ref{fig:ablation_gate_natural}.

The full model consistently outperforms the ungated variant across all candidate-set sizes, and the gain appears in both Top-$K$ accuracy and NBG. At $K=1$, CRS-LLM achieves $70.9\%$ accuracy and $82.0\%$ NBG, compared with $68.9\%$ accuracy and $80.8\%$ NBG without gating. At $K=2$, the gap becomes larger: accuracy increases from $80.2\%$ to $87.4\%$, and NBG increases from $88.0\%$ to $93.0\%$. At $K=3$, the full model reaches $94.1\%$ accuracy and $97.3\%$ NBG, while at $K=5$ it reaches $98.6\%$ accuracy and $99.6\%$ NBG. These gains show that the switch-gated design improves not only the Top-1 result, but also the quality of the whole short list.

This behavior is reasonable because the task contains both smooth continuation and abrupt transitions. Some samples are stable, where the next-step label stays close to the current one and temporal continuity is dominant. Other samples are transition-dominated, where mobility, geometry change, or blockage can make another candidate suddenly become the best one. A single shared prediction head has to use the same decision rule for both cases. The switch-gated design reduces this difficulty by letting one branch focus on stable continuation and the residual flip branch focus on correction during abrupt transitions.

The larger gains at $K=2$ and $K=3$ are especially meaningful. They show that the proposed head does more than sharpen the final Top-1 decision. It also improves the ordering of the most likely candidates near the top of the ranked list. This is exactly what matters in practical beam-management pipelines, where the goal is often to reduce the search burden of later refinement instead of relying only on one hard decision. The NBG gains follow the same trend, which further shows that the better ranking is also physically meaningful.

\subsection{Data Scaling Efficiency in the Few-Shot Regime}

\begin{figure}[h]
    \centering
    \includegraphics[width=\linewidth]{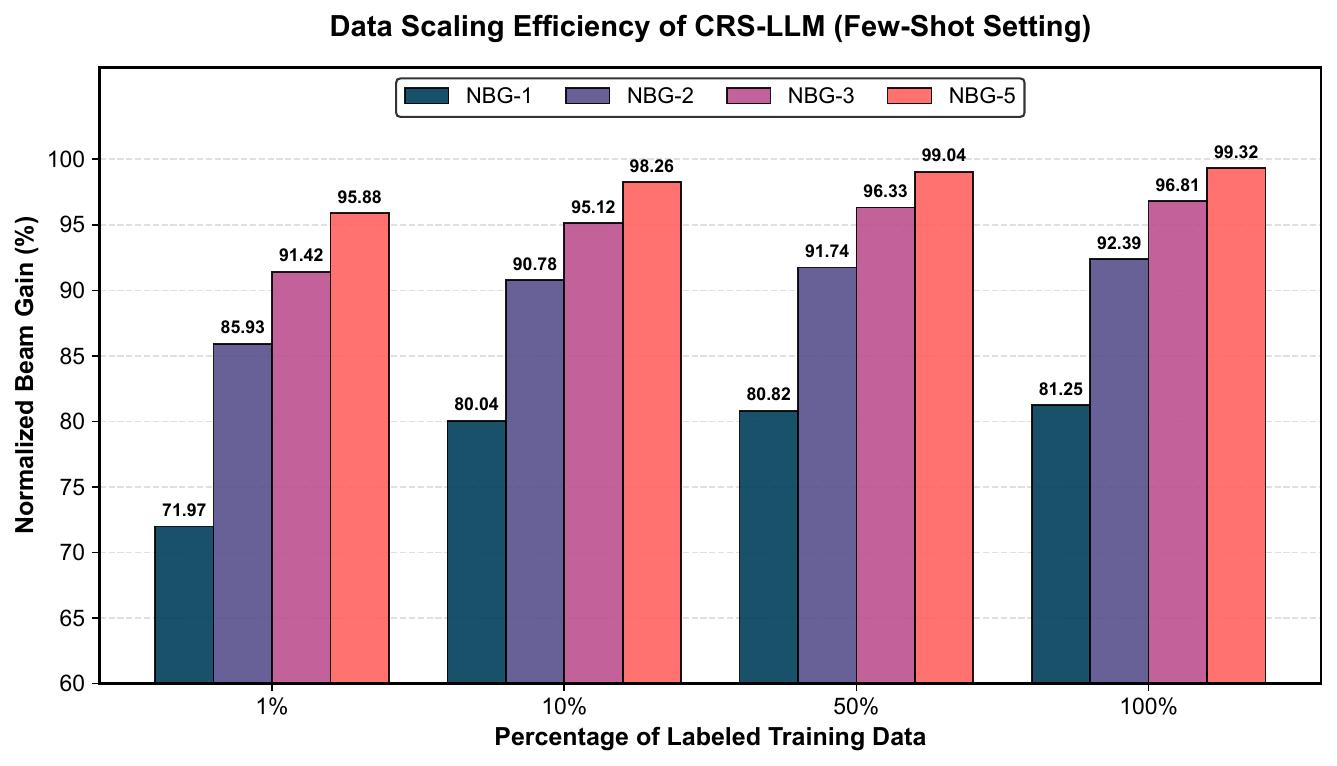}
    \caption{Normalized beam gain under different fractions of labeled training data.}
    \label{fig:data_scaling_natural}
\end{figure}

We next study whether CRS-LLM remains effective when the amount of labeled training data is heavily reduced. The corresponding NBG results under different training fractions are shown in Fig.~\ref{fig:data_scaling_natural}.

A clear observation is that the model remains effective even in the very low-label regime. With only $1\%$ of the labeled training data, CRS-LLM already achieves NBG-1 of $71.97\%$, NBG-2 of $85.93\%$, NBG-3 of $91.42\%$, and NBG-5 of $98.26\%$. When the training fraction increases to $10\%$, the performance further improves to $80.04\%$ for NBG-1 and $99.04\%$ for NBG-5. After that, however, the gain becomes smaller, especially from $50\%$ to $100\%$.

This saturation trend helps explain what the model is learning. If the predictor mainly relied on memorizing narrow label statistics, its performance would be expected to keep scaling more directly with the amount of labeled data. Instead, the strong performance with only a small amount of labels suggests that the model can learn repeated geometric and temporal patterns in cooperative CSI from limited supervision. More labeled data still helps, but its role gradually shifts from learning coarse structure to refining candidate ordering and class boundaries.

Another important point is that the few-shot behavior is especially strong in NBG and larger-$K$ settings. This means that even with limited supervision, the model can still learn a useful ordering of physically plausible BS--beam candidates, which is valuable in practice because a strong short list can already reduce signaling overhead and recovery latency.

\subsection{Regime-Wise Analysis: Stable versus Abrupt-Change Transitions}

\begin{figure}[h]
    \centering
    \includegraphics[width=\linewidth]{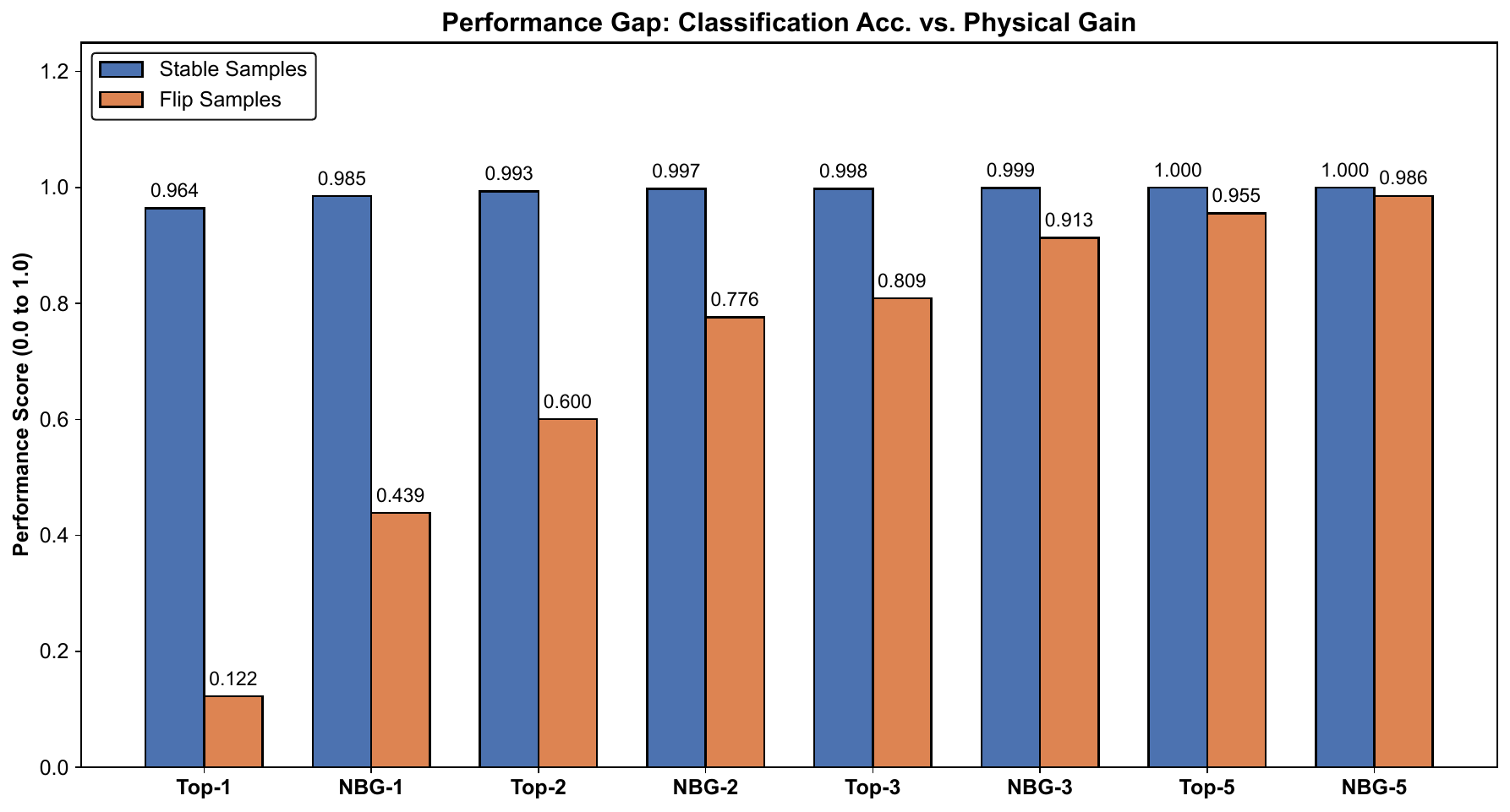}
    \caption{Comparison between stable and abrupt-change regimes in terms of Top-$K$ accuracy and normalized beam gain.}
    \label{fig:stable_flip_natural}
\end{figure}

To better understand where the remaining difficulty lies, we divide the test set into stable and abrupt-change subsets according to whether the future label matches the current label. The corresponding Top-$K$ accuracy and NBG results are shown in Fig.~\ref{fig:stable_flip_natural}.

The stable subset shows highly concentrated next-step prediction. Top-1 accuracy reaches $96.4\%$, and NBG-1 reaches $98.5\%$. As $K$ increases, both Top-$K$ accuracy and NBG quickly approach saturation. This indicates that, for continuation-dominated samples, the model can produce very sharp next-step decisions while also preserving excellent beam-gain quality.

The abrupt-change subset shows a different behavior pattern. Top-1 accuracy is $12.2\%$, and NBG-1 is $43.9\%$. However, performance improves quickly as the candidate-set size grows: Top-2 accuracy and NBG-2 increase to $60.0\%$ and $77.6\%$, Top-3 and NBG-3 rise to $80.9\%$ and $91.3\%$, and Top-5 and NBG-5 reach $95.5\%$ and $98.6\%$, respectively.

These results suggest that, under beam transitions, the model often captures a compact set of plausible next-step candidates even when the final ranking is less concentrated at Top-1. In other words, the difference between the two subsets is not simply whether the model works or fails, but how the prediction quality is distributed between exact Top-1 selection and short-list candidate quality.

This observation also helps explain the role of the switch-gated predictor. The proposed design is useful not only for improving exact continuation-oriented prediction, but also for preserving strong candidate sets when the beam state changes. This is practically meaningful because cooperative beam management often benefits from a compact set of high-quality alternatives rather than one rigid decision alone.

\subsection{Zero-Shot Generalization to Unseen UMa Scenarios}

\begin{figure*}[t]
    \centering
    \begin{minipage}[b]{0.48\textwidth}
        \centering
        \includegraphics[width=\linewidth]{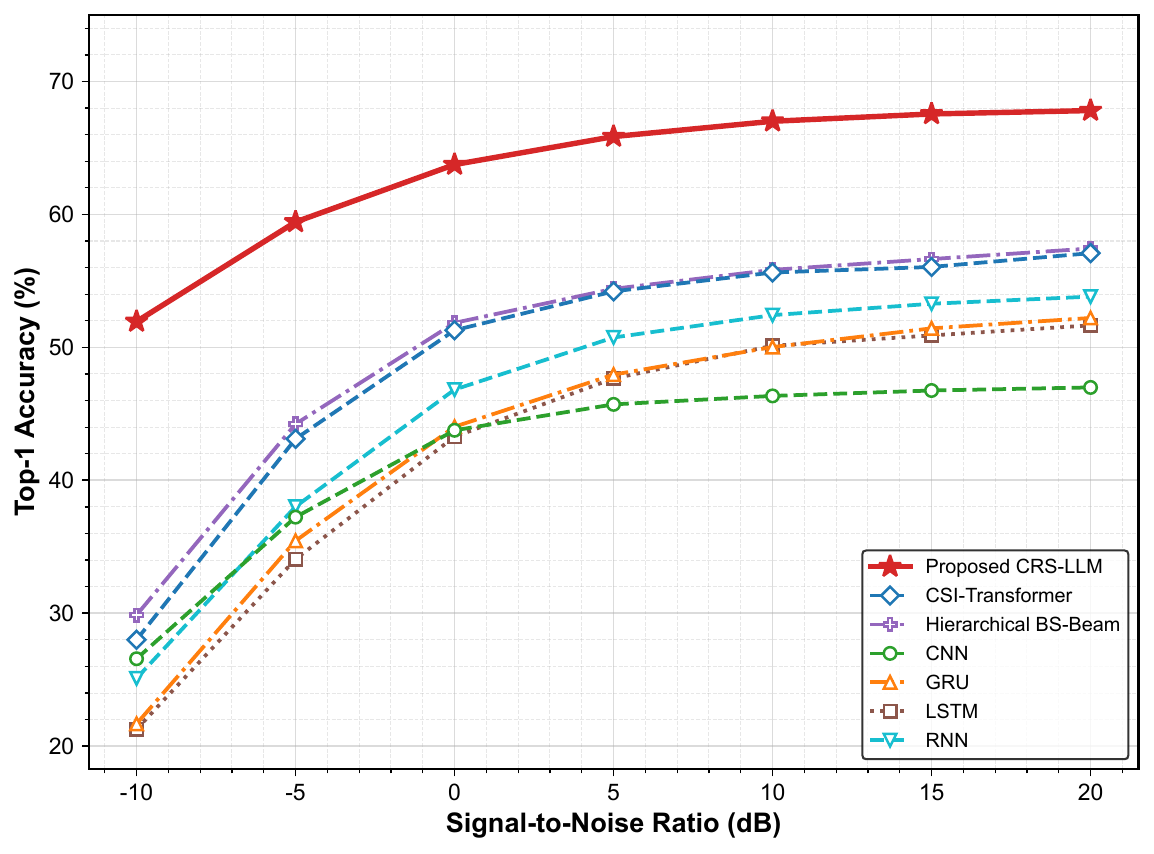}
    \end{minipage}
    \hfill
    \begin{minipage}[b]{0.48\textwidth}
        \centering
        \includegraphics[width=\linewidth]{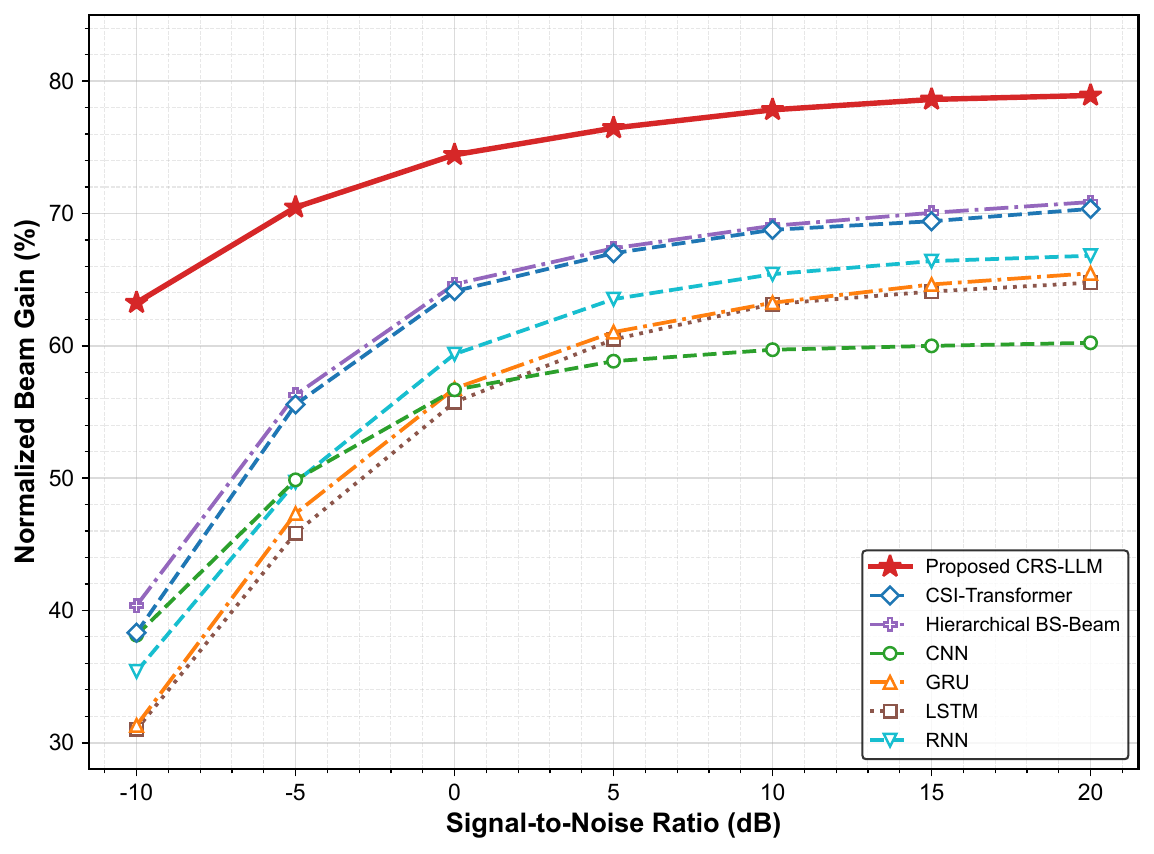}
    \end{minipage}
    \caption{Zero-shot generalization to an unseen UMa scenario: left, Top-1 accuracy; right, NBG.}
    \label{fig:uma_zero_shot_natural}
\end{figure*}

Finally, we examine whether CRS-LLM can generalize beyond the training scenario. To this end, we directly evaluate the trained model in an unseen UMa environment generated under different propagation characteristics, without any additional fine-tuning. The corresponding Top-1 accuracy and NBG results are shown in Fig.~\ref{fig:uma_zero_shot_natural}.

CRS-LLM maintains a clear advantage over all baselines throughout the zero-shot evaluation. Cross-scenario testing is stricter than in-domain SNR testing because it changes not only the observation quality but also the propagation geometry and the beam-evolution statistics. A model that mainly relies on scenario-specific label statistics or simple correlations would usually degrade more strongly under such a shift. The relatively strong performance of CRS-LLM in both Top-1 accuracy and NBG therefore suggests that it learns more transferable structure in cooperative beam dynamics.

In the unseen UMa scenario, CRS-LLM maintains Top-1 accuracy from about $52\%$ to nearly $68\%$ across the full SNR range, while the baselines remain clearly lower. The gap is especially obvious at low SNR, where CRS-LLM stays around $52\%$ at $-10$ dB, while the baselines are only around $22\%\sim30\%$. A similar trend appears in NBG. CRS-LLM preserves about $63\%\sim79\%$ of the optimal beam gain throughout the zero-shot test, while the baselines remain much lower, especially in the more difficult low-SNR region.

In practical terms, CRS-LLM still keeps physically strong candidates near the top of the output ranking even when the exact class is not always predicted correctly in the unseen environment. This robustness is important in deployment, where road layout, blockage density, urban morphology, and BS placement can vary across sites.

\subsection{Overall Discussion}

The results show that CRS-LLM performs consistently well across a wide range of evaluation settings. Its advantage remains visible under both low and high SNR, in optimization behavior, architectural ablation, low-data training, regime-wise analysis, and unseen-scenario transfer. This consistency suggests that the improvement is unlikely to come from accidental tuning or one especially favorable test condition.

The stronger baseline comparisons also help clarify where the gain comes from. The comparison with CSI-Transformer shows that the improvement is not simply due to replacing recurrent models with any transformer-like backbone. The comparison with Hierarchical BS--Beam also supports the value of predicting the joint BS--beam label directly, instead of using a two-stage decision process. These results match the main design choices of CRS-LLM.

The experiments also show that the main parts of the framework play different but complementary roles. The cooperative tokenizer exposes structured CSI patterns and makes the input easier to model, the truncated GPT backbone provides a stronger temporal prior for sequence modeling, and the switch-gated predictor improves robustness across different beam-evolution regimes. Their combination explains why the model shows stable gains across many tests rather than only in one setting.

The regime-wise analysis further shows that CRS-LLM exhibits different strengths across continuation-dominated and transition-dominated conditions. In the former, the model produces highly concentrated Top-1 prediction, while in the latter it still preserves strong Top-$K$ and NBG performance through compact candidate ranking. Overall, CRS-LLM appears well suited to practical cooperative beam prediction systems.

\section{Conclusion}

This paper presented CRS-LLM for cooperative next-step joint BS--beam prediction in multi-BS mmWave/OFDM V2X systems. The proposed framework combines cooperative multi-BS CSI, a dual-view CSI representation, a CNN-based CSI tokenizer, a truncated GPT-style temporal backbone, and a transition-aware switch-gated predictor within a unified joint BS--beam formulation.

Across in-domain, few-shot, and cross-scenario evaluations, CRS-LLM improved both exact prediction and beam-gain recovery relative to the considered baselines. The results further showed that the switch-gated design is particularly useful for preserving strong candidate sets under beam transitions, while the stage-0 warm-up improves optimization stability and the overall framework shows strong performance in low-data settings. Finally, the regime-wise analysis indicates that transition-dominated cases remain the main direction for further improvement.

\bibliographystyle{IEEEtran}
\bibliography{references}

\end{document}